# Optical manipulation of valley coherence via Landau level transitions in black phosphorus and WTe$_2$ monolayers

Xinyu Mu,[1] Shihao Li,[1] Xiaoying Zhou,[2] and Guangyi Jia[1,*]

[1]*School of Science, Tianjin University of Commerce, Tianjin 300134, People's Republic of China*
[2]*Hunan Provincial Key Laboratory of Intelligent Sensors and Advanced Sensor Materials, School of Physics and Electronics Science, Hunan University of Science and Technology, Xiangtan 411201, People's Republic of China*
*Contact author: gyjia87@163.com

Valley coherence is of great significance for exploring fundamental quantum phenomena and developing next-generation valleytronic devices. Herein, we theoretically investigate the valley quantum interference engineered by inter-Landau level (LL) transitions in black phosphorus (BP) and WTe$_2$ monolayers. In contrast to the non-Landau-quantized regime, valley quantum interference is enhanced by over 20-fold, or even significantly stronger, in virtue of striking anisotropic environment. Such anisotropy originates from the distinct electron transition probabilities along the armchair and zigzag directions of BP and WTe$_2$ monolayers. Especially, BP is capable of more effectively strengthening the valley quantum interference response due to its greater directional disparity in electron transition probabilities. The interference fringes also present distinct spectral profiles (e.g., different dip and peak numbers in one interference period) owing to different transition selection rules in BP and WTe$_2$ monolayers. In spite of these discrepancies, normalized interference intensities follow two exponential functions of magnetic field and Landau level index for all the transitions $\delta n = n' - n = 0, \pm 2, \pm 4$ (where $n$ and $n'$ indicate the LL indexes of valence and conduction bands, respectively), and the interference spectra exhibit C$_2$ rotational symmetry about the crystallographic azimuthal angle of 90°.

## I. INTRODUCTION

As an emerging research field, valleytronics has attracted considerable attention for promising applications in next-generation information processing and quantum technologies [1-5]. It focuses on manipulating the valley coherence in two-dimensional (2D) gapped Dirac systems such as transition metal dichalcogenides and gapped bilayer graphene [6-8]. In these materials, the electronic band structures host two degenerate but inequivalent $K$ and $K'$ valleys at the corners of the hexagonal Brillouin zone. The ability to selectively localize particles within these regions of reciprocal space enables the definition of a new binary index characterizing the quantum state of the particle—namely, the valley degree of freedom [9,10]. Importantly, the three-fold rotational symmetry of the Bloch wavefunctions at $K$ and $K'$ gives rise to optical valley selection rules: light of $\sigma_+$ helicity couples only to the $K$ valley, whereas light of $\sigma_-$ helicity couples only to the $K'$ valley [1,11]. This effect allows for valley-dependent interactions between electrons and circularly polarized light, offering a reliable optical strategy for information storage and readout [11,12]. The valley quantum coherence was first identified by the observation of a linearly polarized emission (coherent superposition of $\sigma_\pm$ photons) from monolayer WSe$_2$ optically excited by a linearly polarized light [13]. To date, realizing active control over intervalley quantum coherence has been regarded as a critical step toward enabling realistic applications using valleytronic materials [14-20].

Commonly, an external coherent electromagnetic pump is utilized to generate such quantum coherence between the valleys [13-15]. However, the external coherent pump usually suffers from reliance on complex laser setups, and the intense laser pulses may lead to potential sample damage and strong nonlinear effects [13-15]. Alternatively, valley quantum coherence can be spontaneously generated via creating an anisotropic environment in the vicinity of valley excitons [16-20]. This anisotropic environment can be introduced by placing various metastructures or anisotropic natural crystals near the valleytronic materials [16-20]. Notably, the existing studies employing these metastructures (e.g., plasmonic metasurfaces and moiré superlattices [16-18]) and anisotropic natural crystals (such as $\alpha$-MoO$_3$ and $\beta$-Ga$_2$O$_3$ [19,20]) have primarily focused on tailoring the dielectric environment or polaritonic modes to modulate valley coherence, while completely disregarding the influence of Landau level (LL) transitions on valley coherence.

As discrete energy states, Landau-like levels can be formed by different methods including imposing an external magnetic field and strain-induced pseudomagnetic fields [21-23], which have a great potential for altering the energy distribution and wavefunction interference of valley excitons. The absence of such investigations cleaves a research gap: the underlying mechanism of valley quantum coherence in anisotropic environments has not been fully elucidated without considering LL transitions. The influences of LL indexes, transition selection rules, and transition probabilities on valley coherence remain bewildering. These unresolved issues may hinder the further exploitation of microscopic approaches to precisely manipulating the valley degree of freedom.

In the present work, we theoretically propose an approach to modify the valley quantum interference of two orthogonal valley excitons by magnetizing neighboring monolayer $WTe_2$ or black phosphorus (BP). The discrepancies between valley quantum interferences tuned by $WTe_2$ and BP monolayer as well as the impacts of LL indexes, transition selection rules, transition probabilities, and crystallographic azimuth angles on valley quantum interferences are clearly uncovered. The possible physical mechanism is discussed in detail.

## II. MODEL AND THEORETICAL METHODS

It is assumed that a valleytronic material sits above another anisotropic polaritonic material. The latter is magnetized monolayer crystal of $WTe_2$ or BP, as sketched in Fig. 1(a) or S1(a). The inset in Fig. 1(a) schematically illustrates two degenerate valleys $K$ and $K'$ in the electronic band structure of a hypothetical valleytronic material. Excitons hosted in $K$ and $K'$ valleys are coupled to photons with the helicities $\sigma_{\pm}$, respectively. We consider an initial configuration where one electron is photoexcited to the lowest conduction-band level at the $K$ valley. In an isotropic electromagnetic environment (e.g., free space), this excited electron relaxes to the valence-band maximum by emitting a photon, without inducing any excitation in the orthogonal $K'$ valley. However, in the presence of a neighboring anisotropic material which creates a local in-plane anisotropic space near the valleytronic material, radiative emission from the $K$ valley can resonantly excite an electron in the $K'$ valley, and vice versa. This inter-valley radiative coupling enables the spontaneous emergence of valley coherence and gives rise to quantum interference between the optical emissions from two valleys [19,20].

The electron-hole pairs (i.e., the excitons) in $K$ and $K'$ valleys are modelled as two orthogonal unit electric dipoles which are aligned along $x$- and $y$-axis directions, respectively. Valley coherence can be experimentally quantified using the degree of linear polarization (DoLP) of the optical emissions, defined as DoLP = $(I_1 - I_2)/(I_1 + I_2)$. Here, $I_1$ and $I_2$ represent the intensities of two linearly polarized emissions. The steady-state DoLP is mathematically ill-defined in the absence of external pumping, given that carrier populations in $K$ ($K'$) valleys and the corresponding valley coherence converge to zero [20,24]. To measure the spontaneous valley coherence, a weak incoherent bidirectional pump is assumed. Under this condition, the DoLP is approximately equal to the quantum interference $Q$ which is expressed as [20,24]

$$Q = (\Gamma_1 - \Gamma_2)/(\Gamma_1 + \Gamma_2). \quad (1)$$

Here, $\Gamma_1$ and $\Gamma_2$ denote the spontaneous emission rates for dipoles aligned along $x$- and $y$-axis directions, respectively. The spontaneous emission rate of an arbitrarily oriented emitter situated near an anisotropic material can be obtained from the Purcell factor [25,26]

$$F_p = \frac{\Gamma_{1,2}}{\Gamma_0} = 1 + \frac{6\pi c}{\omega}\left(\boldsymbol{\mu}_{1,2}^* \cdot \mathrm{Im}\left[\overline{\overline{G}}_s(\boldsymbol{r}_0,\boldsymbol{r}_0,\omega)\right] \cdot \boldsymbol{\mu}_{1,2}\right), \quad (2)$$

where $\Gamma_0$ and $c$ are the spontaneous emission rate and the speed of light in vacuum, respectively, $\omega$ is the angular frequency of photons, $\boldsymbol{r}_0$ is the position vector of the dipole, $\boldsymbol{\mu}_{1,2}$ is the unit vector in the direction of the dipole moment of emitter. $\overline{\overline{G}}_s(\boldsymbol{r}_0,\boldsymbol{r}_0,\omega)$ is the scattering part of dyadic Green function at the position of the dipole taking into the presence of anisotropic $WTe_2$ or BP film. Following the procedure outlined in Refs. [25,26], the scattered-tensor Green's function at the source position can be derived by

$$\begin{aligned}&\overline{\overline{G}}_s(\boldsymbol{r}_0,\boldsymbol{r}_0,\omega)\\&=\frac{i}{8\pi^2}\int_{-\infty}^{\infty}\int_{-\infty}^{\infty}\left(r_{ss}\overline{\overline{M}}_{ss}+r_{sp}\overline{\overline{M}}_{sp}+r_{ps}\overline{\overline{M}}_{ps}+r_{pp}\overline{\overline{M}}_{pp}\right)\\&\quad\times\exp(2ik_z z_0)dk_x dk_y\end{aligned} \quad (3)$$

where, the factors $k_x$, $k_y$, and $k_z$ are the components of wave vector along $x$-, $y$-, and $z$-axes, respectively, $z_0$ is the vertical distance of valley excitons from the surface of $WTe_2$ or BP film. Valley excitons are treated as orthogonal dipoles whose position should be sufficiently far from $WTe_2$ or BP film such that they interact with the monolayer film as a whole rather than with individual atoms. Considering the thickness of monolayer $WTe_2$ or BP is about 1 nm, the factor $z_0$ is taken to be 40 nm.

The matrices $\overline{\overline{M}}_{ij}$ ($i, j = s, p$) in Eq. (3) are given by the following expressions [19,25,26]

$$\overline{\overline{M}}_{ss} = \frac{1}{k_z k_\rho^2}\begin{pmatrix} k_y^2 & -k_x k_y & 0 \\ -k_x k_y & k_y^2 & 0 \\ 0 & 0 & 0 \end{pmatrix}, \quad \overline{\overline{M}}_{sp} = \frac{1}{k_0 k_\rho^2}\begin{pmatrix} -k_x k_y & -k_y^2 & -k_y k_\rho^2/k_z \\ k_x^2 & k_x k_y & k_x k_\rho^2/k_z \\ 0 & 0 & 0 \end{pmatrix},$$

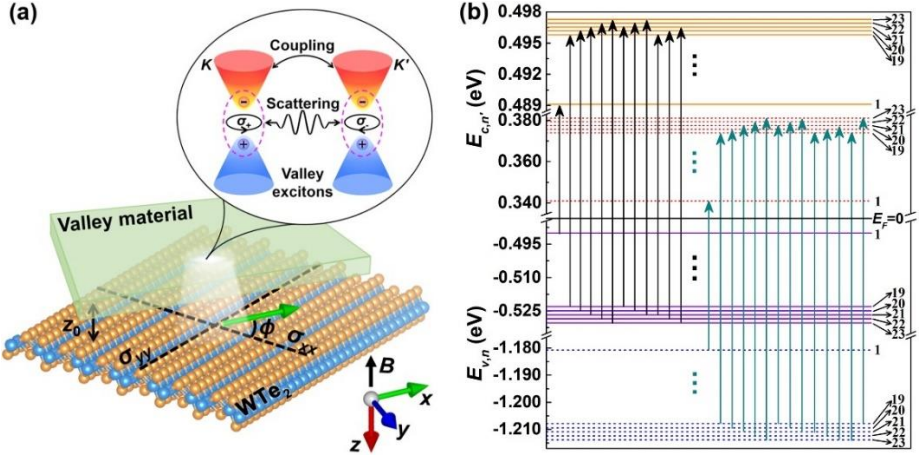

FIG. 1. (a) Schematic depiction of a hypothetical valleytronic material film positioned at a vertical distance $z_0$ from an anisotropic crystal of monolayer WTe$_2$ (This anisotropic crystal can be also monolayer BP, as shown in Fig. S1). Monolayer WTe$_2$ creates a local in-plane anisotropic environment which allows a finite nonzero coupling between mutually orthogonal valley excitons (see the inset) in the valleytronic material film. The radiative emission from $K$ valley with $\sigma_+$ polarization can resonantly excite the orthogonal $K'$ valley with $\sigma_-$ polarization, and vice versa. Such interaction induces spontaneous generation of valley coherence and yields quantum interference among their emissions. (b) At $B = 6$ T, LLs as well as the inter-LL transition rules in monolayer WTe$_2$ (solid horizontal lines, black arrows) and BP (dashed horizontal lines, cyan arrows), where numbers on right side indicate LL indexes. LLs $n(n') \geq 24$ are not sketched, and LLs $2 \leq n(n') \leq 18$ are implied by ellipses.

$$\overline{\overline{M}}_{ps} = \frac{1}{k_0 k_\rho^2}\begin{pmatrix} k_x k_y & -k_x^2 & 0 \\ k_y^2 & -k_x k_y & 0 \\ -k_y k_\rho^2/k_z & k_x k_\rho^2/k_z & 0 \end{pmatrix}, \quad \overline{\overline{M}}_{pp} = \frac{k_z}{k_0^2 k_\rho^2}\begin{pmatrix} -k_x^2 & -k_x k_y & -k_y k_\rho^2/k_z \\ -k_x k_y & -k_y^2 & -k_y k_\rho^2/k_z \\ k_x k_\rho^2/k_z & k_y k_\rho^2/k_z & k_\rho^4/k_z^2 \end{pmatrix}, \qquad (4)$$

where $k_0 = (k_x^2 + k_y^2 + k_z^2)^{1/2}$ is the wave vector in vacuum and $k_\rho = (k_x^2 + k_y^2)^{1/2}$ is the in-plane wave vector.

In Eq. (3), $r_{ij}$ ($i, j = s, p$) are the matrix elements in the tensor of Fresnel reflection coefficient, and they are closely related to complex conductivity tensor $\hat{\sigma}(\omega) = [\sigma_{xx}\ \sigma_{xy};\ \sigma_{yx}\ \sigma_{yy}]$ of WTe$_2$ or BP film. To describe the valley coherence in a general model, monolayer WTe$_2$ or BP is assumed to be suspended in a free space where both the permittivities $\varepsilon_1$ and $\varepsilon_2$ of upper and bottom media are 1.0. We posit monolayer WTe$_2$ or BP is magnetized via a uniform magnetic field $\boldsymbol{B} = -B\hat{z}$, as illustrated in Fig. 1(a). Within the linear-response theory [21,22], the dynamical conductivity components can be written in the usual manner as

$$\sigma_{jk} = i\sigma_0 \frac{\hbar^2}{l_B^2} \times \sum_{n',n,c,v} \frac{[f(E_{c,n'}) - f(E_{v,n})] \langle c,n'|v_j|v,n\rangle \langle v,n|v_k|c,n'\rangle}{(E_{c,n'} - E_{v,n})(E_{c,n'} - E_{v,n} + \hbar\omega + i\Gamma)}, \quad (5)$$

where $j, k \in \{x, y\}$, $\sigma_{xy} = -\sigma_{yx}$ is Hall conductivity, $n$ (or $n'$) is the LL index of valence (or conduction) band, $\sigma_0 = e^2/(\pi\hbar)$, $l_B^2 = \hbar/(eB)$, $f(E_\xi) = \{\exp[(E_\zeta - E_F)/k_B T] + 1\}^{-1}$ is the Fermi-Dirac distribution function with Boltzman constant $k_B$ and temperature $T$; $E_{v,n}$ and $E_{c,n'}$ are the LLs of valence and conduction bands, respectively; and $v_{j/k}$ are the components of group velocities. The sum runs over all states $|\zeta\rangle = |v,n\rangle$ and $|\zeta'\rangle = |c,n'\rangle$ with $\zeta \neq \zeta'$. In this work, the level broadening factor $\Gamma$ and temperature $T$ are set to be 0.15 meV and 5 K, respectively. Besides, we take the Fermi energy $E_F = 0$ such that the contribution from intraband transitions can be ignored [27,28]. Figures S2(a) and S2(b) show the LLs as a function of the magnetic field for the first 50 LLs in monolayer WTe$_2$ and BP, respectively. More details on calculations of LLs and transition matrix elements of the velocity matrices $v_{j/k}$ have been discussed in our previous works [21,22,27,28], and they are reintroduced in section 1 in Ref. [29].

The conductivity component $\sigma_{xx}$ (or $\sigma_{yy}$) is along the armchair (or zigzag) orientation of WTe$_2$ or BP. The $\sigma_{xx}$ axis has a rotation angle $\phi$ to the $x$-$z$ plane in the laboratory Cartesian coordinate. At tunning the rotation angle $\phi$ [see Fig. 1(a)], the anisotropy of monolayer atomic film is characterized via an effective conductivity tensor

$$\hat{\sigma}_{eff} = \begin{pmatrix} \cos\phi & \sin(-\phi) \\ \sin\phi & \cos\phi \end{pmatrix}\begin{pmatrix} \sigma_{xx} & \sigma_{xy} \\ -\sigma_{xy} & \sigma_{yy} \end{pmatrix}\begin{pmatrix} \cos\phi & \sin\phi \\ \sin(-\phi) & \cos\phi \end{pmatrix} = \begin{pmatrix} \sigma_{pp} & \sigma_{ps} \\ \sigma_{sp} & \sigma_{ss} \end{pmatrix}, \quad (6)$$

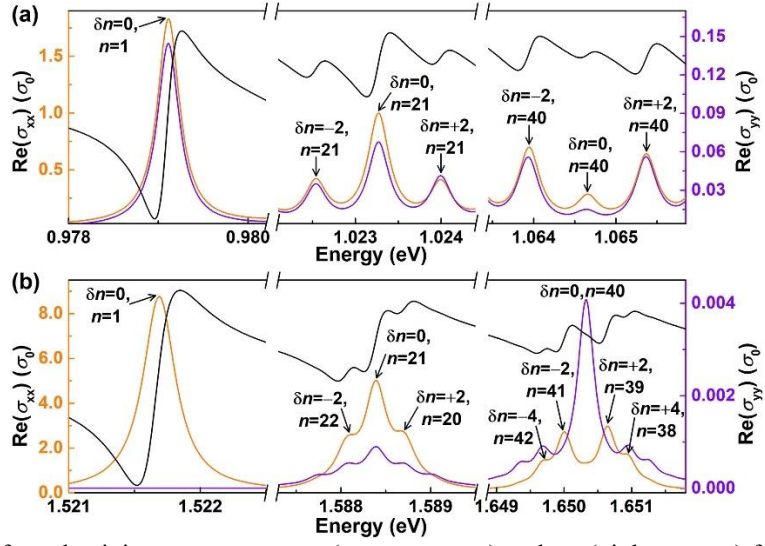

FIG. 2. Real parts of conductivity components $\sigma_{xx}$ (orange curves) and $\sigma_{yy}$ (violet curves) for (a) WTe$_2$ and (b) BP monolayers at $B = 6$ T. Inter-LL transitions $\delta n$ along with the LL index $n$ are determined by matching with LLs in Figs. 1(b) and S2. For comparison, the spectral profiles of $Q$ are plotted by black curves.

where $\sigma_{pp} = \sigma_{xx}\cos^2\phi + \sigma_{yy}\sin^2\phi$, $\sigma_{ss} = \sigma_{xx}\sin^2\phi + \sigma_{yy}\cos^2\phi$, $\sigma_{ps} = (\sigma_{xx}-\sigma_{yy})\sin\phi\cos\phi + \sigma_{xy}$, and $\sigma_{sp} = (\sigma_{xx}-\sigma_{yy})\sin\phi\cos\phi - \sigma_{xy}$. Therefore, the Fresnel reflection coefficients in Eq. (3) are derived via [27,30]

$$r_{pp} = (\beta_+\alpha_-+\chi)/(\beta_+\alpha_++\chi), \quad (7)$$

$$r_{ss} = -(\beta_-\alpha_++\chi)/(\beta_+\alpha_++\chi), \quad (8)$$

$$r_{sp} = -\Lambda(\sigma_{\text{sym}}+\sigma_{xy})/(\beta_+\alpha_++\chi), \quad (9)$$

$$r_{ps} = \Lambda(\sigma_{\text{sym}}-\sigma_{xy})/(\beta_+\alpha_++\chi), \quad (10)$$

where $\alpha_\pm = k_0 \pm k_0 + k_0^2 \sigma_{pp}/(\varepsilon_0\omega)$, $\beta_\pm = k_0 \pm k_0 + \omega\mu_0\sigma_{ss}$, $\chi = (\mu_0\omega)^2(\sigma_{xy}^2-\sigma_{\text{sym}}^2)$, $\Lambda = 2\mu_0\omega^2/c$, $\sigma_{\text{sym}} = (\sigma_{xx}-\sigma_{yy})\sin\phi\cos\phi$, $\varepsilon_0$ and $\mu_0$ are permittivity and permeability in vacuum, respectively.

Hereafter, the rotation angle is set as $\phi = 0°$ unless explicitly specified. After substituting Eqs. (2)-(10) into Eq. (1), the valley quantum interference can be obtained. Because the anisotropic conductivity components, which create a local in-plane anisotropic environment near orthogonal valley excitons, strongly depend on LL transitions, inter-LL transitions in WTe$_2$ and BP films can be used to decisively control over the valley coherence. For simplicity and within the principle of parsimony (also known as the Occam's razor [31]), the valleytronic material is taken to be non-magnetized such that Landau quantization does not occur in valley excitons.

## III. RESULTS AND DISCUSSIONS

Figures S3-S11 (or S12-S20) present the optical spectra of real and imaginary parts of $\sigma_{xx}$, $\sigma_{yy}$ and $\sigma_{xy}$ for monolayer WTe$_2$ (or BP) ranging from 0.977 to 1.160 eV (or 1.515 to 1.780 eV) at $B = 6$, 11, and 16 T. For clarity, Fig. 2 exhibits partially magnified real parts of $\sigma_{xx}$ and $\sigma_{yy}$ spectra of monolayer WTe$_2$ and BP at $B = 6$ T. One can see that all the conductivity spectra are quantized and oscillated with increasing the photon energy. For monolayer WTe$_2$, this multipeak structure in Re($\sigma_{xx,yy}$) gradually transforms from a single-peak to a well-resolved three-peak structure in each oscillation period as the energy increases. By matching with LLs in Fig. S2(a), these three peaks from left to right in Fig. 2(a) correspond to interband transitions when the LL index changes $\delta n = n' - n = -2$, 0, and $+2$. Re($\sigma_{xx}$) is one order of magnitude larger than Re($\sigma_{yy}$) due to the anisotropic band structure in WTe$_2$, and Re($\sigma_{xx,yy}$) are dominated by the transition $\delta n = 0$ (or $\delta n = \pm 2$) in the low (or high) photon energy regime.

Similar multipeak structures are also observed in Re($\sigma_{xx,yy}$) of monolayer BP, as shown in Fig. 2(b). Differing from WTe$_2$, Re($\sigma_{xx}$) is three orders of magnitude larger than Re($\sigma_{yy}$) in BP, indicating a stronger anisotropic environment near valley excitons. According to the interband transition rules in Fig. 1(b), although the three peaks in Fig. 2(a) correspond to different $\delta n$ values, they share with the same LL index $n$ (different $n'$ values). For BP, as exemplified in a moderate energy regime in Fig. 2(b), these three peaks in Re($\sigma_{xx,yy}$) from left to right stem from $\delta n = -2$, 0, and $+2$ with LL indexes $n$ in descending order. Especially, interband transitions $\delta n = \pm 4$ become prominent in the high energy region, resulting in more absorption peaks in Re($\sigma_{xx,yy}$) spectra of BP.

Figure 2 also shows the spectral profiles of $Q$ by black curves. The overall spectra of valley quantum interferences are given in Fig. S22. The interference dips and peaks are mainly distributed on both sides of the transition absorption peak, and the largest $|Q|$ value appears at the energy around the transition $|n=1\rangle \rightarrow |n'=1\rangle$ for all the nonzero $B$ values. For example, the absorption peak of transition $|n=1\rangle \rightarrow |n'=1\rangle$ in WTe$_2$ (BP) appears at 0.9791 eV (1.5217 eV) at $B$ = 6 T while the $Q$ gives its maximum negative and positive values of −0.15 and 0.03 (−0.15 and 0.05) at 0.9789 and 0.9793 eV (1.5215 and 1.5218 eV), respectively. Even if inter-LL transitions create a robust anisotropic dielectric environment, strong interband absorptions are detrimental to photon emissions. As a result of the competition between contributions from anisotropy and optical losses, the dip and peak positions of $Q$ spectra deviate from the transition energy.

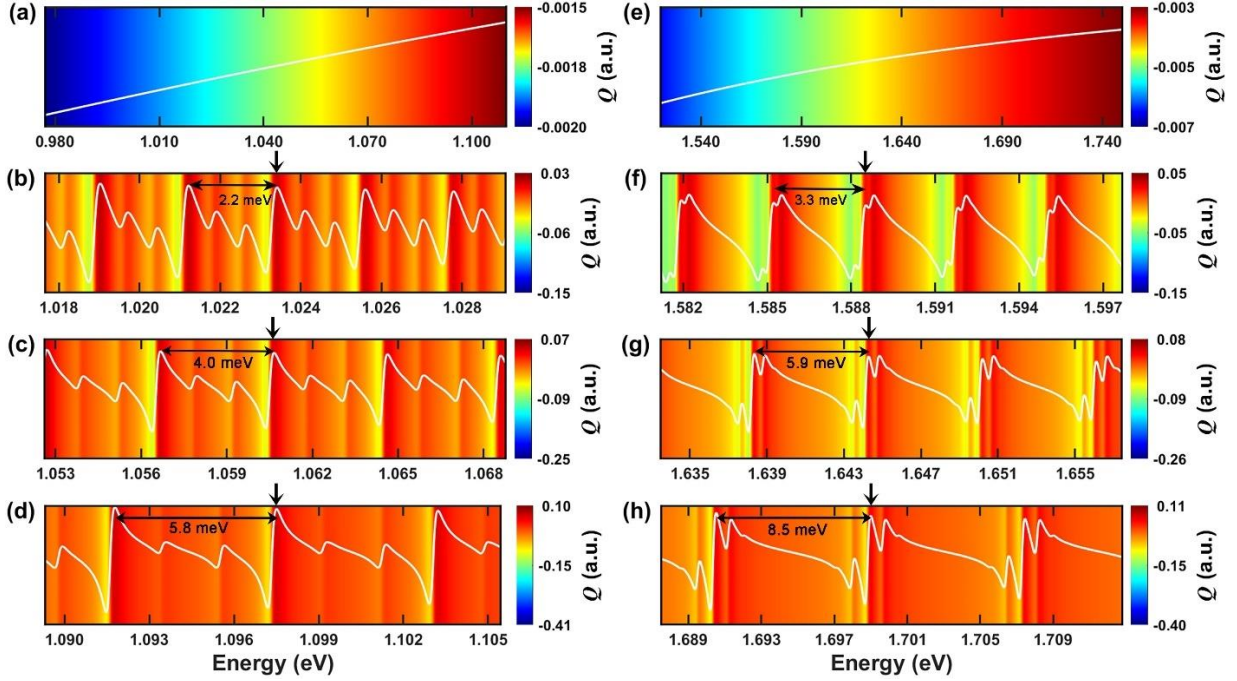

FIG. 3. Quantum interferences $Q$ for two orthogonal dipoles positioned at a vertical distance from (a)-(d) WTe$_2$ and (e)-(h) BP monolayers. (a) and (e), (b) and (f), (c) and (g), (d) and (h) are for the conditions of $B$ = 0, 6, 11, 16 T, respectively. For clarity, the spectral profiles of $Q$ are overlaid as white solid lines in each subgraph. Peak positions of $Q$ nearby the transition $|n=21\rangle \rightarrow |n'=21\rangle$ are marked by vertical one-way arrows. Double arrows indicate the energy separations between interference peaks induced by $|n=20\rangle \rightarrow |n'=20\rangle$ and $|n=21\rangle \rightarrow |n'=21\rangle$.

In the low energy regime, both the conductivities of WTe$_2$ and BP are dominated by the transition $\delta n = 0$ such that only one peak appears in each interference period of $Q$ spectra. As the energy increases, high index LLs are gradually involved in the transition process, and $Q$ spectra in Figs. 2(a) and 2(b) exhibit different peak profiles by virtue of aforementioned different transition rules in WTe$_2$ and BP. Especially for the high energy regime, Re($\sigma_{xx}$) of BP is dominated by the transitions $\delta n = \pm 2$ and $\pm 4$ (i.e., the absorption peak of $\delta n = 0$ disappears due to the weakened influence of interband coupling [28]) such that the $Q$ spectrum presents four interference peaks. By contrast, there are at most three interference peaks in one period of $Q$ spectrum for the case of WTe$_2$.

Figure 3 shows partially magnified $Q$ spectra in Fig. S22. Furthermore, to compare with the unmagnetized condition, optical conductivities of monolayer WTe$_2$ and BP at $B$ = 0 T are also calculated by using the Lorentz-Drude model [32,33]. As shown in Fig. S21 [29], the quantum oscillation disappears and Hall conductivity equals to zero. Moreover, longitudinal and transverse optical conductivities are at the same order of magnitude. Consequently, the spectra of valley quantum interferences display a smooth, featureless energy dependence, consisting with the absence of Landau quantization, and the largest $|Q|$ values for WTe$_2$ and BP are 0.002 and 0.007, respectively, as demonstrated by Figs. 3(a) and 3(e). In marked contrast, the splitting of LLs in WTe$_2$ and BP leads the $Q$ spectra to be quantized and exhibit clear interference fringes. Compared with the unmagnetized conditions in Figs. 3(a) and 3(e), Landau quantization in WTe$_2$ (or BP) by $B$ = 6 T induces the valley quantum interferences to be magnified by 75.0 (or 21.4) times, as presented in Figs. 3(b) and S22(b) [or Figs. 3(f) and S22(f)].

In Fig. 3, peak positions of $Q$ nearby transitions $|n=21\rangle \rightarrow |n'=21\rangle$ are marked by vertical one-way arrows. Meanwhile, the energy separations between interference peaks induced by $|n=20\rangle \rightarrow |n'=20\rangle$ and $|n=21\rangle \rightarrow |n'=21\rangle$ are marked via double arrows. At $B$ = 6 T, the interference peaks are separated by 2.2 and 3.3 meV for the conditions of WTe$_2$ and BP, respectively [see Figs. 3(b) and 3(f)]. Figures 3(c) and 3(d) [or Figs. 3(g) and 3(h)] show that this energy separation increases to 4.0 meV (or 5.9 meV) at 11 T and 5.8 meV (or 8.5 meV) at 16 T for $WTe_2$ (or BP), following a nearly linear scaling with $B$. This behavior directly reflects the LL spacings, where the linearly increased energy separation in $Q$ spectrum indicates the linear broadening of LL spacings induced by the magnetic field.

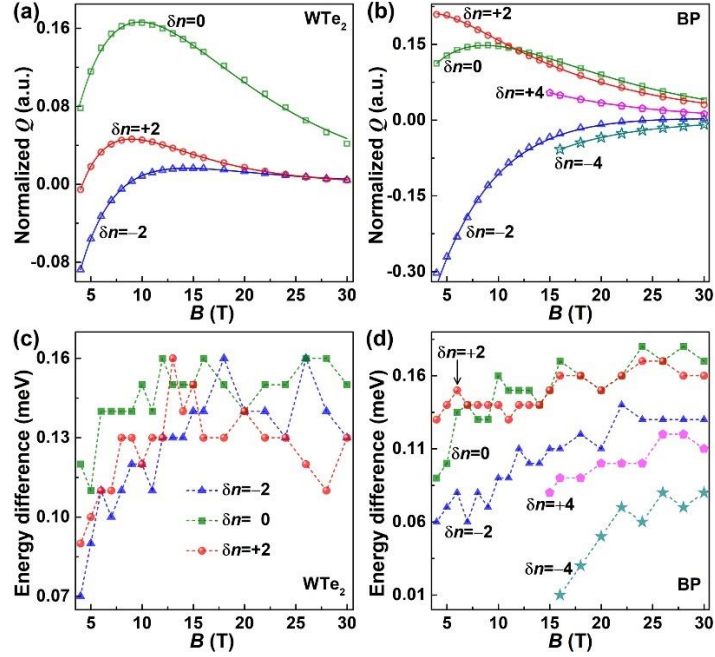

FIG. 4. (a) and (b) Variations of normalized $Q$ peak values with respect to the magnetic induction intensity $B$. Open symbols are the calculated results from Eq. (1). Solid lines are the fitting results by using the empirical Eq. (11). (c) and (d) Energy differences between $Q$ peak and inter-LL transition energies versus the $B$. Symbols are the calculated values while dashed lines are just drawn as a guide to the eye. (a) and (c) are for monolayer $WTe_2$ at $\delta n = 0, \pm 2$ with LL index $n = 21$. (b) and (d) are for monolayer BP at $\delta n = -4, -2, 0, +2,$ and $+4$ with LL indexes $n = 23, 22, 21, 20,$ and $19$, respectively.

To our knowledge, the strongest magnetic field achievable in a personal laboratory is 7 T [34,35]. However, to entirely uncover the evolution of valley quantum interference with the magnetic induction intensity, the largest $B$ value in our theoretical model reaches to 30 T. Additionally, the relative intensity of $Q$ spectra could be more meaningful in experimentally observing Figs. 3 and S22. In view of this, the $Q$ spectra at each magnetic induction intensity are normalized by their respective maximum $|Q|$ values. Figures 4(a) and 4(b) present the normalized $Q$ peak values as a function of $B$ for the cases of $\delta n = 0, \pm 2$ in $WTe_2$ ($n = 21$) and BP ($n = 22, 21,$ and $20$), respectively. When the magnetic field increases to be $\geq 16$ T, the $Q$ spectrum exhibits five-peak structure due to the additional contributions from $\delta n = \pm 4$ ($n = 23$ and $19$) in BP [see Fig. 4(b)].

On the basis of these calculated results, i.e., the open symbols in Figs. 4(a) and 4(b), we construct an empirical formula between normalized $Q$ peak values with the $B$ by

$$Q_{nor}(B) = a_0 \exp(b_0 B) + c_0 \exp(d_0 B), \quad (11)$$

where the factors $a_0, b_0, c_0,$ and $d_0$ are fitting coefficients and their detailed values are listed in Table S1 [29]. The fitting curves in Figs. 4(a) and 4(b) show that $Q_{nor}(B)$ is well in line with the calculated results from Eq. (1). It captures an important trend that normalized $Q$ is an exponential function of $B$.

Figures 5(a) and 5(b) depict the variations of normalized $Q$ peak values at $B = 6$ T with respect to LL index $n$ in $WTe_2$ and BP, respectively. When LL index $n$ increases to be $\geq 37$ in BP, the interference peak contributed by $\delta n = 0$ disappears. This phenomenon is similar to the four-peak structure in the high energy regime of $Q$ spectrum in Fig. 2(b). According to the calculated results, i.e., the open symbols in Figs. 5(a) and 5(b), we find that normalized $Q$ peak values can be well fitted via the following empirical formula

$$Q_{nor}(n) = \sum_{i=1}^{3} a_i \exp\left[-\left(\frac{n-b_i}{c_i}\right)^2\right], \quad (12)$$

where the factors $a_i, b_i, c_i,$ and $d_i$ are fitting coefficients and their detailed values are listed in Table S2 [29].

These fitting coefficients in Tables S1 and S2 are obtained via numerical calculations, verifying the universality of the trend across different LL transitions ($\delta n$

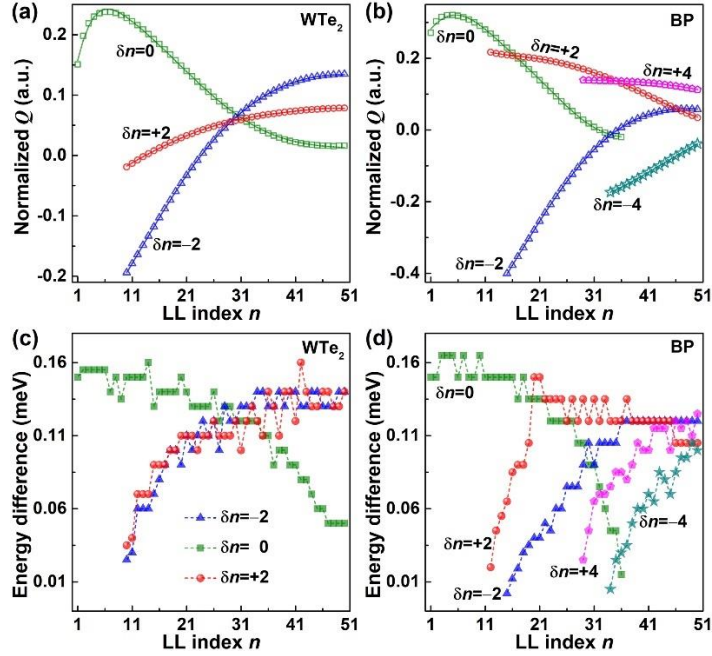

FIG. 5. At the transitions $\delta n = 0, \pm 2$ in monolayer WTe$_2$, (a) normalized $Q$ peak values and (c) energy differences between $Q$ peak and inter-LL transition energies versus LL index $n$. At the transitions $\delta n = 0, \pm 2$, and $\pm 4$ in monolayer BP, (b) normalized $Q$ peak values and (d) energy differences between $Q$ peak and inter-LL transition energies versus LL index $n$. The magnetic induction intensity is $B = 6$ T. In (a) and (b), open symbols are the calculated results from Eq. (1) while solid lines are the fitting results by using the empirical Eq. (12). In (c) and (d), symbols are the calculated values while dashed lines are just drawn as a guide to the eye.

= 0, $\pm 2$, and $\pm 4$). The empirical Eqs. (11) and (12) provides quantitative tools to predict valley quantum interference from magnetic field and LL index. Conversely, LL indexes in 2D materials as well as the external magnetic induction intensity are retrievable from the measurements of valley coherence. Even if Eqs. (11) and (12) have different forms, both of them validate that normalized $Q$ is an exponential function of the factor of $B$ or $n$. This kind of similarity may be rooted in the physical picture: the separation between LLs in the valence band and conduction band can be increased by either enhancing the magnetic field or increasing the LL index $n$ ($n'$).

As mentioned above, the dips or peaks of interference fringes are deviated from the transition absorption peak of Re($\sigma_{xx,yy}$). Figures 4(c) and 4(d) present the energy differences between $Q$ peak and inter-LL transition energies at different $B$ values. With the increase of magnetic induction intensity, the energy difference exhibits quantum oscillations locally, while the overall tendency first increases and then tends to stabilize for all the transitions $\delta n = 0, \pm 2$, and $\pm 4$. The similar variation tendency is replicated at the transitions $\delta n = \pm 2$ and $\pm 4$ via increasing LL index $n$ at a fixed magnetic field, as shown in Figs. 5(c) and 5(d). The possible reason is that increasing the magnetic field enhances the intensities of absorption peaks originating from transitions $\delta n = 0, \pm 2, \pm 4$, as exemplified by partially magnified Re($\sigma_{xx}$) spectra in the insets in Figs. S3 and S12. At a fixed magnetic field, the absorption intensities of transitions $\delta n = \pm 2, \pm 4$ gradually increase with the LL index $n$, as demonstrated by Re($\sigma_{xx,yy}$) spectra in Fig. 2. To "escape" such absorption losses, the interference peaks shift away from the transition energy as the magnetic field or the LL index $n$ increases. Nonetheless, the absorption peak at $\delta n = 0$ gradually weakens as the LL index $n$ increases (see Fig. 2), thus, the overall tendency of energy difference between $Q$ peak and inter-LL transition energies at $\delta n = 0$ is decreasing with increasing LL index $n$, as shown by green square symbols in Figs. 5(c) and 5(d). Note that Re($\sigma_{yy}$) is three orders of magnitude smaller than Re($\sigma_{xx}$) in BP [see Fig. 2(b)]. Therefore, the variation tendency of energy difference is dominated by Re($\sigma_{xx}$) although the absorption peak of Re($\sigma_{yy}$) at $\delta n = 0$ in BP gradually increases as the LL index $n$ increases.

The above results show that valley coherence can be manipulated by BP in a larger range than that modulated via WTe$_2$. Superficially, this reflects that BP possesses stronger anisotropy than WTe$_2$. The underlying physical mechanism originates from different transition probabilities which are proportional to the squares of the matrix elements of velocity matrices [28,29,36]. According to Eqs. (S11) and (S12), the anisotropic magneto-optical conductivity of WTe$_2$ mainly arises from the ratios $v_{f1}/v_{f2} = 3.56$ and $v_1/v_2 =$ 1.28 [29]. In monolayer BP, the Fermi velocity $v_f$ = 7.95×10$^5$ m/s, the velocities along armchair and zigzag orientations are $v_1$ = 3.96$B^{1/2}$×10$^3$ m/s and $v_2$ = 3.83$B^{1/2}$×10$^3$ m/s, respectively [29]. Based on Eqs. (S15) and (S16) [29], the anisotropic magneto-optical conductivity of BP mainly emanates from the ratio $v_f/v_2$ which is two orders of magnitude larger than the velocity ratios in WTe$_2$. This is consistent with the results in Figs. 2 and S3-S20 that the ratios Re($\sigma_{xx}$)/Re($\sigma_{yy}$) and Im($\sigma_{xx}$)/Im($\sigma_{yy}$) of BP are ~10$^2$ larger than that of WTe$_2$. In consequence, the stronger valley coherence modulated by BP than that by WTe$_2$ arises essentially from the larger discrepancy between electron transition probabilities along armchair and zigzag orientations in BP compared with that in WTe$_2$.

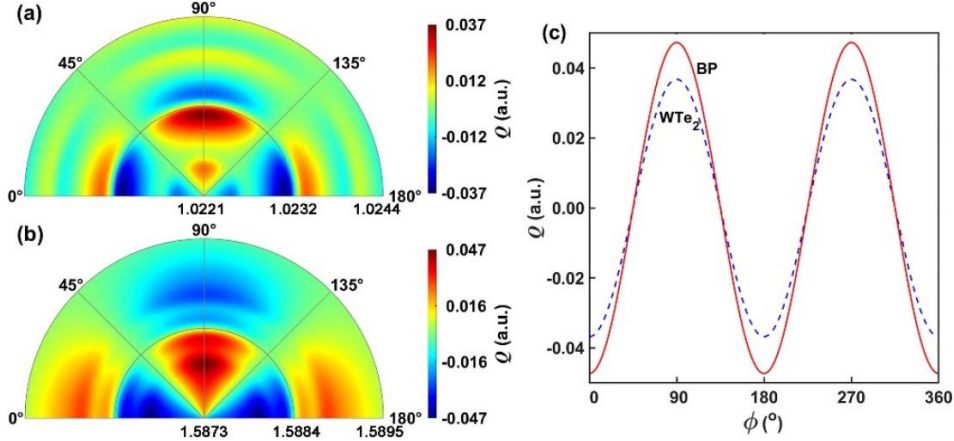

FIG. 6. For the cases of (a) WTe$_2$ and (b) BP, variations of quantum interference $Q$ with respect to the photon energy $\hbar\omega$ and the rotation angle $\phi$. The radial axis and angle stand for the factors of $\hbar\omega$ in eV and $\phi$ in degrees, respectively. (c) $Q$ values versus the rotation angle $\phi$ at $\hbar\omega$ = 1.0231 eV for WTe$_2$ and $\hbar\omega$ =1.5880 eV for BP. The magnetic induction intensity is $B$ = 6 T.

Finally, the impact of rotation angle $\phi$ on valley quantum interference is also inspected, as presented in Fig. 6. The photon energy range is around the transition $|n=21\rangle \rightarrow |n'=21\rangle$. One can see that the interference spectrum exhibit C$_2$ rotational symmetry about the 90° axis, with $Q$ distributions mirroring across this angular line. The interference has the largest $|Q|$ value at $\phi = 0°$ and 90°. Besides, owing to its larger directional disparity in electron transition probabilities, BP can be better to amplify the valley quantum interference response in comparison with WTe$_2$.

## IV. CONCLUSIONS

In summary, valley coherence engineered by inter-LL transitions in BP and WTe$_2$ monolayers has been investigated. In comparison with the unmagnetized condition, valley quantum interference $Q$ can be magnified to more than 20 times (or even larger) by virtue of the anisotropic dielectric environment induced via Landau quantization of energy bands in BP and WTe$_2$ monolayers. This anisotropy derives from the difference in electron transition probabilities along armchair and zigzag orientations of 2D atomic crystals. The interference fringes are a direct reflection of the LL spacings, and the interference spectra modulated by WTe$_2$ and BP exhibit distinct spectral profiles (e.g., different dip and peak numbers in one interference period) as a result of different transition selection rules. For normalized interference intensities, they follow two different exponential functions of the magnetic induction intensity and LL index for all the transitions $\delta n$ = 0, ±2, and ±4. Through rotating the azimuth angle of WTe$_2$ and BP, the interference spectrum exhibit C$_2$ rotational symmetry about the 90° axis. Relative to WTe$_2$, BP is capable of more effectively enhancing the valley quantum interference response due to its greater directional disparity in electron transition probabilities. These predicted physical phenomena as well as the uncovered interplay mechanism between valley excitons and LLs in time-reversal symmetry breaking 2D quantum systems provide vistas and theoretical support for the microcosmic design of next-generation active valleytronic quantum devices.

## ACKNOWLEDGMENTS

The authors acknowledge the financial support by the National Natural Science Foundation of China (Grant No. 11804251).

## DATA AVAILABILITY

The data that support the findings of this article are not publicly available. The data are available from the authors upon reasonable request.

[1] F. Gucci, E. B. Molinero, M. Russo, P. San-Jose, F. V. A. Camargo, M. Maiuri, M. Ivanov, A. Jiménez-Galán, R. E. F. Silva, S. Dal Conte, and G. Cerullo, Encoding and manipulating ultrafast coherent valleytronic information with lightwaves, Nat. Photonics **20**, 266–272 (2026).

[2] Y. Oda, M. P. Losert, and J. P. Kestner, Suppressing Si valley excitation and valley-induced spin dephasing for long-distance shuttling, Phys. Rev. Lett. **136**, 020802 (2026).

[3] H. Jiang, Y. Zhang, L. H. An, Q. H. Tan, X. R. Dai, Y. Z. Chen, W. J. Chen, H. B. Cai, J. T. Fu, J. Zúñiga-Pérez, Z. W. Li, J. H. Teng, Y. Chen, C. W. Qiu, and W. B. Gao, Chiral light detection with centrosymmetric-metamaterial-assisted valleytronics, Nat. Mater. **24**, 861–867 (2025).

[4] R. Xu, Z. G. Zhang, J. Liang, and H. Y. Zhu, Valleytronics: Fundamental challenges and materials beyond transition metal chalcogenides, Small **21**, 2402139 (2025).

[5] K. Mourzidis, V. Jindal, M. Glazov, A. Balocchi, C. Robert, D. Lagarde, P. Renucci, L. Lombez, T. Taniguchi, K. Watanabe, T. Amand, S. Francoeur, and X. Marie, Exciton formation in two-dimensional semiconductors, Phys. Rev. X **15**, 031078 (2025).

[6] H. N. Wang, K. Shinokita, K. Watanabe, T. Taniguchi, S. Konabe, and K. Matsuda, Direct identification of valley coherence and its manipulation in monolayer two-dimensional semiconductor, ACS Nano **19**, 21484–21491 (2025).

[7] Z. W. Yan, H. Ma, Y. J. Zhu, X. L. Zhang, R. X. Bai, X. S. Du, R. Zhou, Y. S. Tang, R. C. Mao, K. Watanabe, T. Taniguchi, C. Sevik, and C. Y. Jiang, Unhybridized interlayer excitons enabled by heterostrain and electric field in bilayer $WSe_2$, ACS Nano **20**, 2707–2716 (2026).

[8] T. Xie, S. Y. Xu, Z. Y. Dong, Z. Y. Cui, Y. B. Ou, M. Erdi, K. Watanabe, T. Taniguchi, S. A. Tongay, L. S. Levitov, and C. H. Jin, Long-lived isospin excitations in magic-angle twisted bilayer graphene, Nature **633**, 77–82 (2024).

[9] O. Huber, K. Kuhlbrodt, E. Anderson, W. Li, K. Watanabe, T. Taniguchi, M. Kroner, X. Xu, A. Imamoglu, and T. Smolenski, Optical control over topological Chern number in moiré materials, Nature **649**, 1153–1158 (2026).

[10] M. Kim, T. Kim, A. Galler, D. Kim, A. Chacon, X. X. Gong, Y. H. Yang, R. L. Fang, K. Watanabe, T. Taniguchi, B. J. Kim, S. H. Chae, M. H. Jo, A. Rubio, O. Neufeld, and J. Kim, Quantum interference and occupation control in high harmonic generation from monolayer $WS_2$, Nat. Commun. **16**, 9825 (2025).

[11] H. Huang, B. Tian, Y. Chen, X. Y. Xia, X. M. Cui, L. Shao, H. J. Chen, and J. F. Wang, Coherent polaritons in $WSe_2$-monolayer-sandwiched Au-nanodisk-on-mirror structures, ACS Nano **19**, 25284–25294 (2025).

[12] I. Tyulnev, Á. Jiménez-Galán, J. Poborska, L. Vamos, P. S. J. Russell, F. Tani, O. Smirnova, M. Ivanov, R. E. F. Silva, and J. Biegert, Valleytronics in bulk $MoS_2$ with a topologic optical field, Nature **628**, 746–751 (2024).

[13] A. M. Jones, H. Y. Yu, N. J. Ghimire, S. F. Wu, G. Aivazian, J. S. Ross, B. Zhao, J. Q. Yan, D. G. Mandrus, D. Xiao, W. Yao, and X. D. Xu, Optical generation of excitonic valley coherence in monolayer $WSe_2$, Nat. Nanotech. **8**, 634–638 (2013).

[14] Y. Shi, Y. S. Gan, Y. Z. Chen, Y. B. Wang, S. Ghosh, A. Kavokin, and Q. H. Xiong, Coherent optical spin Hall transport for polaritonics at room temperature, Nat. Mater. **24**, 56–62 (2025).

[15] M. Wegerhoff, M. Scharfstädt, S. Linden, and A. Bergschneider, Coherent interaction of 2s and 1s exciton states in transition-metal dichalcogenide monolayers, Phys. Rev. Lett. **134**, 236901 (2025).

[16] P. K. Jha, N. Shitrit, X. X. Ren, Y. Wang, and X. Zhang, Spontaneous exciton valley coherence in transition metal dichalcogenide monolayers interfaced with an anisotropic metasurface, Phys. Rev. Lett. **121**, 116102 (2018).

[17] Y. Zhu, K. L. Zou, D. X. Qi, J. He, R. W. Peng, and M. Wang, Tailoring valley polarization of interlayer excitons in van der waals heterostructure toward optical communication, Nano Lett. **25**, 8680–8688 (2025).

[18] X. Xie, B. Wu, J. N. Ding, S. F. Li, J. Y. Chen, J. He, Z. W. Liu, J. T. Wang, and Y. P. Liu, Emergence of optical anisotropy in moire superlattice via heterointerface engineering, Nano Lett. **24**, 9186–

9194 (2024).

[19] A. Bapat, S. Dixit, Y. Gupta, T. Low, and A. Kumar, Gate tunable light-matter interaction in natural biaxial hyperbolic van der Waals heterostructures, Nanophotonics **11**, 2329–2340 (2022).

[20] G. Y. Jia, J. X. Luo, C. Y. Cui, R. J. Kou, Y. L. Tian, and M. Schubert, Valley quantum interference modulated by hyperbolic shear polaritons, Phys. Rev. B **109**, 155417 (2024).

[21] Q. Y. Ma, H. Dou, Y. T. Chen, G. Y. Jia, and X. X. Zhou, Photonic spin Hall effect dependent on Landau level transitions in monolayer $WTe_2$, Phys. Rev. B **113**, 075420 (2026).

[22] G. Y. Jia, Q. Z. Cai, C. Q. Zheng, X. Y. Zhou, and C. W. Qiu, Canalized hyperbolic magnetoexciton polaritons enabled by the Shubnikov-de Haas effect in van der Waals semiconductors, Phys. Rev. B **113**, 075407 (2026).

[23] J. Guo, G. Q. Xu, M. Q. Liu, X. Zhou, G. M. Tao, and C. W. Qiu, Pseudo-Landau thermal diffusion, Phys. Rev. Lett. **136**, 056306 (2026).

[24] M. Nalabothula, P. K. Jha, T. Low, and A. Kumar, Engineering valley quantum interference in anisotropic van der Waals heterostructures, Phys. Rev. B **102**, 045416 (2020).

[25] L. Novotny and B. Hecht, *Principles of Nano-Optics* (Cambridge University Press, Cambridge, 2012).

[26] B. Sikder, S. H. Nayem, and S. Z. Uddin, Deep ultraviolet spontaneous emission enhanced by layer dependent black phosphorus plasmonics, Opt. Express **30**, 47152–47167 (2022).

[27] G. Y. Jia, G. Li, Y. Zhou, X. L. Miao, and X. Y. Zhou, Landau quantisation of photonic spin Hall effect in monolayer black phosphorus, Nanophotonics **9**, 225–233 (2020).

[28] X. Y. Zhou, W. K. Lou, F. Zhai, and K. Chang, Anomalous magneto-optical response of black phosphorus thin films, Phys. Rev. B **92**, 165405 (2015).

[29] See Supplemental Material at http://link.aps.org/supplemental/XXXX for calculations on LLs and transition matrix elements, complex optical conductivities of monolayer $WTe_2$ and BP at $B$ = 0, 6, 11, and 16 T, Figs. S1−S22, and tables S1−S2.

[30] W. J. M. Kort-Kamp, B. Amorim, G. Bastos, F. A. Pinheiro, F. S. S. Rosa, N. M. R. Peres, and C. Farina, Active magneto-optical control of spontaneous emission in graphene, Phys. Rev. B **92**, 205415 (2015).

[31] E. Fischer, Guiding principles in physics, Eur. J. Philos. Sci. **14**, 65 (2024).

[32] C. Wang, S. Y. Huang, Q. X. Xing, Y. G. Xie, C. Y. Song, F. J. Wang, and H. G. Yan, Van der Waals thin films of $WTe_2$ for natural hyperbolic plasmonic surfaces, Nat. Commun. **11**, 1158 (2020).

[33] A. Nemilentsau, T. Low, and G. Hanson, Anisotropic 2D materials for tunable hyperbolic plasmonics, Phys. Rev. Lett. **116**, 066804 (2016).

[34] L. Wehmeier, S. H. Xu, R. A. Mayer, B. Vermilyea, M. Tsuneto, M. Dapolito, R. Pu, Z. Y. Du, X. Z. Chen, W. J. Zheng, R. Jing, Z. J. Zhou, K. Watanabe, T. Taniguchi, A. Gozar, Q. Li, A. B. Kuzmenko, G. L. Carr, X. Du, M. M. Fogler, D. N. Basov, and M. K. Liu, Landau-phonon polaritons in Dirac heterostructures, Sci. Adv. **10**, eadp3487 (2024).

[35] R. A. Mayer, X. Z. Chen, R. Jing, M. Tsuneto, B. Y. Zhou, Z. J. Zhou, W. J. Zheng, R. Pu, S. H. Xu, T. Liu, H. L. Yao, L. Wehmeier, Y. N. Dong, D. H. Sun, L. He, A. R. Cadore, T. Heinz, J. A. Fan, C. R. Dean, D. N. Basov, X. Du, R. O. Freitas, and M. K. Liu, Magnetically tunable polariton cavities in van der waals heterostructures, Nano Lett. **25**, 13079–13086 (2025).

[36] V. Aji, Adler-Bell-Jackiw anomaly in Weyl semimetals: Application to pyrochlore iridates, Phys. Rev. B **85**, 241101 (2012).